\documentclass[12pt]{article}
\usepackage{a4}
\title{Chromo-electric Yang-Mills gauge fields}
\author{Vinod Chandra\thanks{email: vinodc@iitk.ac.in}, Ravindra Kumar\thanks{email: rojha@iitk.ac.in}
\\Department of Physics,\\
Indian Institute of Technology, Kanpur, India\\
 Kanpur-208016}
\date{\today}
\begin{document}
\maketitle
\begin{abstract}
We obtain static gauge field configurations for $SU(2)$ Yang-Mills gauge theory in the absence of chromo-magnetic field. We also present a systematic way to obtain such configurations for $SU(3)$ and $SU(4)$ YM gauge theories by realizing their Lie algebras in terms of $SU(2)$. Generalization of our method to $SU(N)$ gauge theory with N$>2$ is straight forward. The gauge fields thus obtained are complex. We investigated the issue of uniform as well as real chromo-electric field with these configurations. Furthermore, we  investigated the possible class of vacuum solutions for the gauge fields and found that they are gauge inequivalent to the usual vacuum configuration $A^a_{\mu}=0$.
\end{abstract}

\vspace{2mm}
The classical solution of Yang-Mills(YM) equations plays central role to get physical insights about the vacuum as well as non-perturbative nature  of the theory about which much is still unknown. Earlier, there were many  attempts to solve  YM equations for the classical gauge field configurations. It is worth to mention a few of them. A very first step is due to Wu and Yang\cite{wu1}, in which they proposed static sourceless solutions for YM equations. Later on, YM theories in the presence of external sources have been studied in a great detail in the following works\cite{ch,sw,jm,jackw}. Interestingly, real as well as complex solutions for the YM gauge fields have been found. In the work by Villarroel\cite{villa}, in which  a class of complex configurations for the Yang-Mills gauge fields for $SU(2)$ gauge theory in the Minkoswki space time has been obtained and their physical relevance is highlighted. Later on, Hoker and Vinet\cite{hv} discussed the complex solutions for the gauge theories in detail. Such configurations has also been reached as a solution of (2+1) dimensional Yang-Mills equations with the chern-simon term\cite{osr}. These solutions are of great interest because of the fact that, due to a general theorem of Wu and Yang \cite{wu1,wu2}, the complex gauge fields of the group $SU(2)$ can be converted to the real solutions of $SL(2,C)$.

 In this paper, we shall not solve YM equations rather, we attempt to obtain those configurations which give zero chromo-magnetic field\footnote{The word chromo is used to distinguish the electrodynamics and non-abelian gauge theories and it stand for color of non-abelian gague group. Since, we  nowhere refer electrodynamics hence we shall omit the word chromo henceforth}(in the rest Lorentz frame) as well as uniform electric field by considering an ansatz. Such configurations could be useful to study transport phenomena and particle production in heavy ion collisions.

The plan of this paper is as follows. We initiate our study with $SU(2)$ YM gauge theory and obtain these configurations for the gauge fields. Furthermore, we present a systematic way to obtain them for $SU(3)$ and $SU(4)$ YM gauge theories by realizing their Lie structure in terms of interlinked $SU(2)$. This approach may easily be generalized to $SU(N)$ with N$>2$. Having known all the spatial component $A^a_i$, we shall indicate how to obtain $A^a_0$ so as to get uniform electric field. In particular, we obtain $A^a_0$ in the case of zero electric field. Interesting point is to mention that these configurations for the gauge fields are complex and gauge inequivalent to $A^a_{\mu}$.

{\bf The  YM gauge fields for zero magnetic field:}\\
Recall that, the field strength tensor for $SU(N)$ YM gauge theory reads,
\begin{equation}
F^a_{\mu\nu}=\partial_\mu A^a_{\nu} - \partial_\nu A^a_{\mu} + g f^{abc} A_{\mu}^b A_{\nu}^c
\end{equation}
where, a, b ,c are the color indices and runs from $1$ to $(N^2-1)$ for $SU(N)$, g is the coupling strength and $f^{abc}$ are the structure constants of the Lie algebra of $SU(N)$. The electric and magnetic fields are defined in the same manner as in the case of electrodynamics with a additional color index as follows,\\
$E_i ^{a}=F^{a}_{0i}$ and $B^{a}_i=\frac{\epsilon^{ijk}}{2} F^{a} _{jk}$ . Zero magnetic field $(B^{a}=0)$ leads to the following equation,
\begin{eqnarray}
\label{eqn1}
\epsilon^{ijk}\bigg[\partial_j A^a_{k} - \partial_k A^a_{j} + g f^{abc} A_{j}^b A_{k}^c\bigg]=0 \nonumber\\
\Longrightarrow \nabla\times \vec A^a = -\frac{g}{2} f^{abc} \vec A^b \times \vec A^c \nonumber\\
\end{eqnarray}
The above equation(eq.\ref{eqn1}) is a set of $N^2-1$ number of self consistent equations. As a next step, we shall consider $SU(2)$ gauge group and obtain the solution of eq.\ref{eqn1} and thereafter we shall systematically obtain them in the case of  $SU(3)$ and $SU(4)$ respectively.

 {\bf SU(2) gauge theory:}
Let us consider $SU(2)$ gauge group. Recall that, for $SU(2)$  $f^{abc} \equiv \epsilon^{abc}$, where $\epsilon^{abc}$ is levi-cevita tensor. We consider an ansatz $A^1=-i\nabla\phi$ to solve eq.\ref{eqn1}. With this ansatz one obtains the follwing configurations for the gauge fields ,
\begin{eqnarray}
\label{eqn3}
\phi = \log(y+z),  \vec A^{1}= i \frac{(-g^{-1})}{y+z}(\hat{j}+\hat{k}) \nonumber\\
\vec A^{2}=-i\frac{1}{y+z}\hat{ j}, \vec A^{3}=\frac{1}{y+z}\hat{j}\nonumber\\
\end{eqnarray}
 We found that the above configurations belong to the following general class  of ansatz  in which $\vec A^2$ is parallel to $\vec A^3$ in some gauge, displayed below,
\begin{equation}
\label{eqn2}
\vec A^1=\nabla\phi, \vec A^2=i\exp(i\phi)\nabla\psi, \vec A^3=\exp(i\phi)\nabla\psi
\end{equation}
with $\phi=-i\log(y+z)$ and $\nabla\psi=\hat{j} 1/(y+z)^{2}$. Note that, for simplicity we put $g=1$ in eq.\ref{eqn1}.

An important point here is to be noted that the gauge field configurations displayed in eq.\ref{eqn3} are
complex. Recall that complex gauge field configurations for $SU(2)$ are well studies in the Euclidean space  \cite{vsg} and also in the Minkowski space time\cite{villa}. They pointed out that the complex solutions are physically acceptable as far as they give real energy density. We shall come back to the physical importance of them in the later part of this paper. Here, we wish further to obtain such configurations for $SU(3)$ and $SU(4)$ YM theories. The idea is to look at the structure of their algebras in relation to $SU(2)$.

{\bf $SU(3)$ and $SU(4)$ gauge theories:}
   It is well known that the structure of the lie algebra of $SU(3)$ generators, allows us to decompose it into three interlinked $SU(2)$. Out of the eight generators of $SU(3)$, which are none other than the Gellmann matrices $\lambda$'s. The algebra's of the three interlinked $SU(2)$, can be written as,
$
 [\lambda_1, \lambda_2]=2i\lambda_3,
 [\lambda_4, \lambda_5]=2i(\frac{\lambda_3 +\sqrt(3)\lambda_8}{2}),
 [\lambda_6, \lambda_7]=2i(\frac{\lambda_3 -\sqrt(3)\lambda_8}{2}).$ Corresponding to each of these $SU(2)$ subalgebras, we have three
equations analogous to eq.\ref{eqn1} and use the ansatz same as in the previous part of this paper to obtain zero magnetic field configurations. Further, these configurations will not be independent of one another and we have freedom to choose, $\vec A^1=\vec A^4=\vec A^6$,$\vec A^5=(1-s)\vec A^2$,$\vec A^7=(1+s)\vec A^2$, $\vec A^8=s\frac{1}{\sqrt{3}}\vec A^3$, where s is an arbitrary constant. The explicit forms may easily be written by utilizing configurations displayed in eqs.\ref{eqn3}, which reads,\\
$\vec A^1= \vec A^4=\vec A^6=i \frac{(-g^{-1})}{y+z}(\hat{j}+\hat{k})$\\
$\vec A^5=(1-s)-i\frac{1}{y+z}\hat{ j}$, $\vec A^7=(1+s)-i\frac{1}{y+z}\hat{ j}$ and\\ $\vec A^8=s\frac{1}{\sqrt{3}(y+z)}\hat j$.

With the same philosophy as for $SU(3)$, one may also obtain complex solution for SU(4). It is straight forward to see that the first eight generators $(\lambda_1 - \lambda_8)$ of $SU(4)$ form a 4-dimensional $SU(3)$ and the rest of the generators form three additional $SU(2)$ as,

$[ \lambda_9, \lambda_{10}]=2i(\frac{\lambda_3}{2}+\frac{\lambda_8}{2\sqrt{3}}+\sqrt{\frac{3}{2}} \lambda_{15})$,
$[\lambda_{11}, \lambda_{12}]=2i(-\frac{\lambda_3}{2}+\frac{\lambda_8}{2\sqrt{3}}+\sqrt{\frac{3}{2}} \lambda_{15})$, and
$\bigg[\lambda_{13}, \lambda_{14}\bigg]=2i(-\frac{\lambda_8}{\sqrt{3}}+\sqrt{\frac{3}{2}} \lambda_{15})$.

 We obtain the following solutions for $SU(4)$ Yang-Mills gauge fields,
$\vec A^1=\vec A^4= \vec A^6=\vec A^9=\vec A^{11}=\vec A^{13}$,
$\vec A^2=(1+s)^{-1}\vec A^5=(1-s)^{-1}\vec A^7$,
$\vec A^8=\frac{s \vec A^3} {\sqrt{3}}$,
$\vec A^{10}=(1+s1)(1+s)\vec A^2$,
$\vec A^{12}=((1-s1)-s(1+s1))\vec A^2$, and
$ \vec A^{15}=\frac{s1(1+s)}{2}\sqrt{\frac{3}{2}}\vec A^3.$
Note that s and s1 are arbitrary constants. Again the explicit form may be obtained by substituting the configurations displayed in eqs. \ref{eqn3} and \ref{eqn2}.
  This approach can easily be generalized to obtain such configurations for $SU(N)$ with $N>2$. For the details of $SU(N)$ Lie algebra, one may see the following ref.\cite{wp}.

 {\bf The uniform electric field:}
Let us come to the physical relevance of such configurations.
Recall that, in the colour flux tube models the issue of quark-confinement is assumed to be implemented by the generation of electric flux tube with uniform energy density\cite{cash}. The uniform energy density and the string tension in such approaches is related to the uniform electric field, which has potential applications to study particle production\cite{cash,vr1} and space-time evolution\cite{vr1,vr2} of quark-gluon-plasma at RHIC and LHC.  We, therefore, wish to address the question that what kind of static gauge field configurations give uniform electric field.  Recall that, electrostatic field can be given by,
\begin{equation}
\label{el}
  E^a_i=-\partial_i  A^a_0 +g f^{abc}  A^{b} _0 A^{c} _i
\end{equation}
where $A^a_0$ is the scalar potentials. So far, the form of $A^a_0$ has not been determined. It is easy to see that  the form is not independent since it is related with $A^a$ by the YM equations. Note that the form of $A^a_0$ is determined by demanding various physical situations of interest. In our case, it has been determined by solving eq.\ref{el}  by considering uniform electric field. 
 One may find real electric fields by utilizing the gaugue field configurations for $SU(2)$ displayed in eq.\ref{eqn2}, with the gaugue choice $A_0^a=i\delta^a_3$. The electric field configuration thus obtained are displayed below,\\
$\vec E^1=\hat j \frac{1}{y+z}$, $\vec E^2=\frac{g^{-1}}{y+z}(\hat j + \hat k)$ and 
$\vec E^3= \vec 0$. Important point here is to be noted that , although the gague field  are complex but the the above electric field configurations corresponds to real and finite energy-density, hence physically acceptable. The uniform electric field  can be found by the choice $A_0^a=i(y+z)\delta^a_3$. In this case $\vec E^1$ and $\vec E^2$ are real on the other hand $\vec E^3$ is purely imaginary. Nevertheless, the uniform electric field configurations also give real energy density. To obtain such configurations for $SU(3)$ and $SU(4)$ is straight forward. One may obtain them by the technique mentioned in the previous part this paper.

 {\bf Zero energy configurations:}
 The zero energy gauge field configurations may be classified as,
(i) $A^a_0=0$, $\vec A^a=0$; (ii) $A^a_0=0$, $\vec A^a\ne 0$;(iii) $A^a_0\ne 0$, $\vec A^a\ne 0$. Having obtained the form of  $A^a-i$ (eq.\ref{eqn2}), the scalar potential $A^a_0$ can be obtained  by solving the follwing equation,
\begin{equation}
\label{eq6}
\nabla A^a_0 -g f^{abc} A^{b} _0 \vec A^{c} =0.
\end{equation}
For $SU(2)$ gauge theory, we obtain,
$A^2_0\sim cos(\int^x A^1_x dx + \int^y A^1_y dy+ \int^z A^1_z dz)$, $A^3_0\sim sin(\int^x A^1_x dx + \int^y A^1_y dy+ \int^z A^1_z dz)$ and $A^1_0$ is given in terms of the condition,\\  
$\nabla A^1_0\sim A^2(\exp(-i(\int^x A^1_x dx + \int^y A^1_y dy+ \int^z A^1_z dz))$.
Note that $A^a_0=0$ trivially satisfies eq.\ref{eq6}. Substitution of gague field configurations displayed in  eq.\ref{eqn2} leads following configurations for $A^a_0$,\\
 $A_0^1\sim (constt.), A_0^2\sim cos(2i \ln(y+z) + k), \\
 A_0^3 \sim sin(2i \ln(y+z) +k)$.

Where k is a constant.

The most interesting observation regarding to the vaccum configurations is the following: 
the zero energy gauge field configurations which falls in classes  (ii) and (iii) are gauge inequivalent to the usual vacuum  configurations given by class (i). To get such configurations for $SU(N)$ with N $> 2$, it is again useful to exploit the idea developed in the earlier part of this paper.

 {\bf Conclusions and comments:}
We have explicitly obtained complex static configurations of the gauge fields for  $SU(2)$ Yang-Mills theory in the absence of magnetic field. We have presented a systematic way to obtain such configurations for $SU(3)$ and $SU(4)$ Yang-Mills theories by realizing their Lie structures in terms of interlinked $SU(2)$ sub-algebras. This approach can  easily be generalized to $SU(N)$ gauge theories with N$>2$. We discussed the case of uniform chromo-electric field and obtained the configurations for $A^a_0$. These configurations are needed  for studying transport phenomena and particle production in heavy ion collisions. We shall address this in the future work.

An special attention has been paid to zero energy solution. It has been found that configuration obtained
are gauge inequivalent to the vacuum configuration given by $A_{\mu}^a$=0. We wish to make a remark that  make that our ansatz given by eq.\ref{eqn2} in which $\vec A^2$ is parallel to $\vec A^3$ is not a
gauge independent statement.  As a final remark, we wish to state that although, we have only considered time independent case, we have freedom to take time dependence in the configurations given by
eq.\ref{eqn3} and \ref{eqn2}. The time dependence of $A^a_0$ may be reached by solving $E^a_i=-\partial _t A^a_i-\partial_i A^a_0 +g f^{abc} A^{b} _0 A^{c} _i$ for $A^a_0$. The quantum corrections to these time dependent configurations may be useful for the study of some of the physical phenomena of interest.

\noindent
We are grateful to Prof V. Ravishankar for the useful discussions . We also  acknowledge CSIR(India) for the financial support through the award of a fellowship.

\end{document}